\documentclass[10pt,letterpaper]{article}
\usepackage[top=0.85in,left=2.75in,footskip=0.75in]{geometry}
\usepackage{amsmath,amssymb}
\usepackage{changepage}
\usepackage[utf8x]{inputenc}
\usepackage{textcomp,marvosym}
\usepackage{cite}
\usepackage{nameref,hyperref}
\usepackage{microtype}
\DisableLigatures[f]{encoding = *, family = * }
\usepackage{float}
\usepackage[table]{xcolor}
\usepackage{ulem}
\usepackage{array}
\newcolumntype{+}{!{\vrule width 2pt}}
\newlength\savedwidth

\raggedright
\usepackage[aboveskip=1pt,labelfont=bf,labelsep=period,justification=raggedright,singlelinecheck=off]{caption}

\makeatletter
\renewcommand{\@biblabel}[1]{\quad#1.}
\makeatother
\usepackage{lastpage,fancyhdr,graphicx}
\usepackage{epstopdf}
\fancyheadoffset[L]{2.25in}
\fancyfootoffset[L]{2.25in}
\begin{document}
\vspace*{0.2in}
\begin{flushleft}
{\Large\textbf\newline{Response of the competitive balance model to the external field} }
\newline
\\
Farideh Oloomi\textsuperscript{1},
Amir Kargaran\textsuperscript{1},
Ali Hosseiny\textsuperscript{1},
Gholamreza Jafari\textsuperscript{1*,2}
\\
\bigskip
\textbf{1} Department of Physics, Shahid Beheshti University, Tehran, Iran
\\
\textbf{2} Irkutsk National Research Technical University, Irkutsk, Russia

\bigskip
* gjafari@gmail.com

\end{flushleft}
\section{Abstract}
		The competitive balance model was proposed as an extension of the structural balance theory, aiming to account for heterogeneities observed in real-world networks. In this model, different paradigms lead to form different friendship and enmity. As an example, friendship or enmity between countries can have a political or religious basis. The suggested Hamiltonian is symmetrical between paradigms. Our analyses show that a balanced state can be achieved if just one paradigm prevails in the network and the paradigm shift is possible only by imposing an external field. In this paper, we investigate the influence of the external field on the evolution of the network. We drive the mean-field solutions of the model and verify the accuracy of our analytical solutions by performing Monte-Carlo simulations.
We observe that the external field breaks the symmetry of the system. The response of the system to this external field,  contingent upon temperature, can be either paramagnetic or ferromagnetic. We observed a hysteresis behavior in the ferromagnetic regime. Once communities are formed based on a certain paradigm, then they resist change. We found that to avoid wasting energy we need to know the level of stochastic behavior in the network. Analogous to magnetic systems, we observe that susceptibility adheres to Curie's law. 


\section{Introduction}
Structural balance theory is a sociological model for describing the dynamics of friendship and enmity in signed networks. This theory considers the presence of 3-cycles in the network and emphasizes the influence of third parties who have connections with both individuals involved in a dyadic relation.The model was originally introduced by Heider in 1946 \cite{heider1946attitudes}. In recent years, the model has become popular to address a wide range of phenomena ranging from interpersonal relationships in society \cite{doreian2004evolution, de2013polarization}, Online interactions through the Web \cite{leskovec2010predicting, leskovec2010signed} to the formation of international coalitions  \cite{hart1974symmetry, lerner2016structural, belaza2017statistical, doreian2019structural}.

 Over the years, numerous studies have proposed modifications to Heider's original formulation of balance theory, and expanded its applicability \cite{kulakowski2005heider, saeedian2017epidemic, gorski2017destructive, singh2016competing, kirkley2019balance, belaza2019social, bagherikalhor2021heider,Masoumi2022Erdos}.
In this regard, the Competitive balance model \cite{oloomi2021competitive} has been proposed as an extension to the balance model to account for the heterogeneity observed in relationships. To this end, this model includes two different types of friendship and enmity. As an example, on the international level, coalitions can form based on political interests or religious ones. 
Previous study has examined the dynamics of this model under symmetrical condition and in the absence of the external forces \cite{oloomi2021competitive}.
Howeve, in the real world, this equality is not always maintained and the symmetry is often broken by the application of external fields. In this paper, we study the dynamics of the competitive balance model in the presence of an external field. We aim to explore the effects of this external force on the formation and evolution of relationships within the network. By analyzing the interplay between the competitive balance model and the external field, we seek to enhance our understanding of the complex dynamics that shape real-world networks.

In the structural balance model, the relationship between members is either friendly or enmity, which is denoted with +1 and -1, respectively. Considering these two types of links, four types of triangles can form. A triangle is considered balanced if the product of its edges is positive, otherwise, it is unbalanced. The mathematical framework of balance theory for collections of more than three members was formulated by Cartwright and Harary \cite{cartwright1956structural}. They showed that a fully connected network is balanced when all the triangles within it are balanced. In 2005, Antal et al. presented two discrete dynamical models to examine how an unbalanced network evolves toward a balanced state \cite{antal2005dynamics}. At the same time, Ku\l akowski et al. formulated continuous dynamic equations for the networks with continuous values of friendship and enmity \cite{kulakowski2005heider}. 

Despite the success of balance theory to describe a wide range of phenomena, we still know that the real-world networks exhibit additional complexities \cite{chakrabarti2006econophysics,gatti2008emergent,niu2021information,battiston2016complexity}. In recent years, increasing attention have been devoted to heterogeneities of real-world networks \cite{gorski2017destructive,singh2016competing,gorski2020homophily} .
It is simplistic to assume that all relationships either friendship or enmity have the same origin. Various influential factors such as religion, politics, and culture can lead to the formation of different kinds of friendships and enmities. As an example, Middle Eastern countries may have religious-based relations among themselves while maintaining politically and economically driven relationships with Western countries. Thus, various factors are influential in the formation of coalitions. The competitive balance model aims to enrich balance theory to cover such heterogeneity and highlight conflicts of interest in socio-political networks \cite{oloomi2021competitive}. In this model, two different interests form two different types of friendships and enmities. In fact, it is not enough to say that the two agents are friends or enemies. You also need to specify the type of friendship or enmity based on the interests of formation. 

The competitive balance model employs real and imaginary numbers to distinguish between different types of relations. In other words, dyadic relations besides $\pm 1$ can get values $ \mp i$ that stand for friendship or enmity based on different interests. The proposed Hamiltonian is symmetrical regarding both types of relations (real and imaginary). Also, the Hamiltonian is reduced to the Hamiltonian of the original balance theory when all links in the network are either real or imaginary. 

The dynamics of such networks have been explored in previous study \cite{oloomi2021competitive}. It has been shown, while the edges switch their type (using Monte Carlo simulation) to minimize the energy, the symmetry is spontaneously broken and ultimately, only one type of link (real or imaginary) prevails in the network. The role of temperature in the model has also been studied \cite{masoumi2021mean}, demonstrating a first-order phase transition from a homogenous ordered phase to a disordered phase.

In this paper, we investigate the impact of external forces on the competitive balance model.  We analyze the system's response to external forces that favor one of the two types. In multi-state systems, the application of an external field breaks the symmetry of the system, leading to a preferred minimum among several possibilities.
The Ising model, which describes the behavior of interacting spins on a lattice, is a well-known multi-state model. Its rich history and established principles make it a proper framework for obtaining valuable insights in our research. The effect of external forces on the Ising model has been extensively studied \cite{sides1998kinetic, korniss2000dynamic, sides1999kinetic}, revealing different regimes of system response depending on the strength of the external field, particularly below the critical temperature. \cite{rikvold1994metastable}. 
Hysteresis properties of the Ising model is studied in Ref. \cite{chakrabarti1999dynamic,misra1997spin} , providing insights into the economic responses of the European Union and the United States to government economic stimuli during the 2009 recession \cite{hosseiny2016metastable, hosseiny2019hysteresis, bahrami2020optimization}.

So, we examine the hysteresis properties of our model and show that the final state of the network depends on the initial condition, temperature, and how the field is applied to it, that is, whether the field is increasing or decreasing during the application.
To find stable configurations in this model, we employ two approaches:  1- mean-field approximation\cite{rabbani2019mean,park2004statistical, park2004solution, park2005solution, kargaran2020quartic, masoumi2021mean}, 2- Monte-Carlo simulation.
Finally, we compare the result of these two approaches, which exhibit good agreement with each other.

\section{Materials and methods}
\subsection{Model}
In balance theory, each pair of agents are either friend or enemy which are labeled by $+1$ and $-1$ respectively. Within this framework, the energy of each triangle is defined as:
\begin{eqnarray}\label{eq1}
-\sigma_{ij}\sigma_{jk}\sigma_{ki},
\end{eqnarray}
where $\sigma_{ij}$ stands for the relation between agents $i$ and $j$. Then, the Hamiltonian of the network is given by
\begin{eqnarray}\label{eq2}
\mathcal{H}_0= -\sum_{i>j>k}\sigma_{ij}\sigma_{jk}\sigma_{ki}.
\end{eqnarray}

In the framework of competitive balance theory, friendships and enmities are categorized into two distinct types based on different underlying foundations. This distinction gives rise to two forms of friendship and two forms of enmity within the relationships. To capture these complex forms of relationship, complex numbers have been employed. In the pair-wise relations, friendship is labeled either by $+1$ or $-i$ which stands for friendships based on the first interest or the second one. Similarly, enmity has two different labels denoted by $-1$ and $+i$. With this in mind, the energy of each triangle can be defined as:
\begin{eqnarray}\label{eq3}
-Re(\sigma_{ij}\sigma_{jk}\sigma_{ki})-Im(\sigma_{ij}\sigma_{jk}\sigma_{ki}),
\end{eqnarray}
where $Re$ indicates the real part of the product and $Im$ indicates the imaginary part. This definition allows us to quantify the energy of each triangle in the network. Interestingly, similar to the structural balance model, the energy of each triangle in the competitive balance model is either $-1$ or $+1$. This energy shows if the triple relation is in tension or not.
So the Hamiltonian of this network in the competitive balance model is given by:
\begin{eqnarray}\label{eq4}
\mathcal{H}_0 = -Re(\sum_{i>j>k}{\sigma_{ij}\sigma_{jk}\sigma_{ki}})-Im(\sum_{i>j>k}{\sigma_{ij}\sigma_{jk}\sigma_{ki}}),
\end{eqnarray}
This definition guarantees symmetry between both forms of relation. Additionally, it is worth noting that when all links in the network are either real or imaginary, the energy of the network reduces to that of the regular balance model \cite{oloomi2021competitive}.

Evolution in the competitive balance model has been studied in Refs \cite{oloomi2021competitive}. In these analyses, an ensemble of fully connected networks initiates the evolution with random initial conditions, where each link is randomly assigned a number from $\pm1,\;\pm i$. Then, using the Monte Carlo method, the edges are randomly selected and updated to minimize the energy as defined in Eq.~\ref{eq4}.  It is observed that during this evolution, the symmetry is spontaneously broken,  leading to the dominance of one form of relation. Consequently, homogeneity emerges within the network with the majority of links being either real or imaginary. 

In this work, we examine the effect of an external field on the dynamic of competitive balance theory. The external field is assumed to exert an impact on the network, favoring one of the interests. To define interaction with the external field, we aim to impose the following restrictions:
\begin{enumerate}
	\item The external field is applied to the links ($\sigma_{ij}$) of the network, not the triangles ($\sigma_{ij}\sigma_{jk} \sigma_{ki}$). 
	\item The external field supports either real or imaginary forms of relation, regardless of whether the relationship is classified as friendship or enmity. 
	\item To ensure that the Hamiltonian remains a real value, we use the quadratic form of  ${\sigma_{ij}}$.
\end{enumerate}

To satisfy the above-mentioned conditions, we add a term to Hamiltonian as
\begin{eqnarray}\label{eq5}
\mathcal{H} = \mathcal{H}_0-h\sum_{i>j}{\sigma^2_{ij}},
\end{eqnarray}
where $h$ represents the external field. If the value of $h$ is positive, then the field term in Eq.(5) i.e. $-h\sum_{i>j}{\sigma^2_{ij}}$ is reduced when links turn real. Conversely, if the value of $h$ is negative, then the field term decreases as the number of imaginary links increases. So, the external field breaks the symmetry in favor of one form of relations.

In the absence of an external field the only term for total energy comes from the energy of triangles which we call “mean-triangle-energy” $E_{\triangle}$. This energy can be expressed as follows:

\begin{eqnarray}\label{eq6}
E_{\triangle}=-\frac{Re(\sum{\sigma_{ij}\sigma_{jk}\sigma_{ki}})+Im(\sum{\sigma_{ij}\sigma_{jk}\sigma_{ki}})}{\mathcal{N}_{\triangle}},
\end{eqnarray}
where $\mathcal{N}_{\triangle}$ is the total number of triangles in the network. This parameter indicates how close the network is to the balanced state. The value of $E_{\triangle}$ varies between $-1$ and $+1$. It is $-1$ when the network is in a balanced state and all triangles are balanced. As the number of unbalanced triangles increases, the value of $E_{\triangle}$ grows, i.e., the network moves away from the balanced state. 

In the absence of an external field, the Hamiltonian is symmetrical and the value of energy can not identify whether the real links dominates or the imaginary links. The second term in Hamiltonian Eq. \ref{eq5}, however, breaks the symmetry and allows us to identify the dominant type of links. 
Since below the critical temperature for any given energy, the system has two different symmetrical equilibrium states, we borrow the concept of magnetization from the field of electromagnetism and call the following parameter the  “generalized magnetization” $M$: 
\begin{eqnarray}\label{eq7}
M = \frac{1}{\mathcal{L}}\sum_{i>j}{\sigma^2_{ij}}= \frac{L_{re}-L_{im}}{\mathcal{L}},
\end{eqnarray}
where $L_{re}$ and $L_{im}$ are respectively the number of real and imaginary links and $\mathcal{L} = (L_{re}+L_{im})$ is the total number of the links in the network. The value of $M$ ranges between -1 and 1, providing insight into the dominance of one form of relations over the other.

In the following sections, we obtain the stable states of our model using the mean-field method. We know that the statistical features of systems are influenced by the dimension of the network \cite{henkel1999conformal, christiansen2020aging, saberi2019competing, dashti2017bak, masoudi2002statistical, goodarzinick2018robustness, najafi2020geometry}. But in the mean field approximation, each element in the system interacts with all other elements in an average or mean-field manner, neglecting spatial correlations or the specific network structure. As a result, the behavior of the system in the mean field universality class is independent of the network size. Fully connected networks belong to the mean-field universality class. So, in our work, we will examine our mean-field solution with the simulation in a fully connected network.

\subsection{\label{sec3}Mean-field solution}

We consider a fully connected network with $N=50$ nodes. Though the solution does not depend on the size, just the quantitative values such as the critical temperature depend on the size.

To start, we separate the share of $\sigma_{ij}$ from the rest of the Hamiltonian, i.e.,
\begin{equation}\label{eq8}
\mathcal{H}=\mathcal{H}_{ij}+\mathcal{H}',
\end{equation}
in which $H_{ij}$ is the sum of all terms in Hamiltonian that contain $\sigma_{ij}$ and $\mathcal{H}'$ includes the remaining terms. So
\begin{equation}\label{eq9}
-\mathcal{H}_{ij}= Re \left[\sigma_{ij} \sum_{k \neq i,j} \sigma_{jk}\sigma_{ki} \right] +Im \left[\sigma_{ij} \sum_{k \neq i,j} \sigma_{jk}\sigma_{ki} \right]+h\sigma_{ij}^2.
\end{equation}

Now, we can calculate the average of physical quantities using the probability distribution $P(G)={e^{-\beta \mathcal{H}}}/{Z}$. Note that $\beta=1/T$ and G denotes the configuration of network and, $Z=\sum_{G}e^{-\beta \mathcal{H}}$ is the partition function. So, the mean value of $\sigma_{ij}$ is:
\begin{equation}\label{eq10}
\begin{aligned}
\langle\sigma_{ij}\rangle&=\frac{1}{Z}\sum_{G}\sigma_{ij}e^{-\beta\mathcal{H}(G)}=\frac{1}{Z}\sum_{\{\sigma\neq\sigma_{ij}\}}e^{-\beta\mathcal{H}'}\sum_{\{\sigma_{ij}=\pm{1}, \mp{i}\}}\sigma_{ij} e^{-\beta\mathcal{ H}_{ij}}\\
&=\frac{\sum_{\{\sigma\neq\sigma_{ij}\}}e^{-\beta\mathcal{H}'}\sum_{\{\sigma_{ij}=\pm 1,\pm\mathrm{\textit{i}}\}}\sigma_{ij}e^{-\beta\mathcal{H}_{ij}}}{\sum_{\{\sigma\neq\sigma_{ij}\}}e^{-\beta\mathcal{H}'}\sum_{\{\sigma_{ij}=\pm 1,\pm\mathrm{\textit{i}}\}}e^{-\beta\mathcal{H}_{ij}}}={\frac{\big\langle\sum_{\{\sigma_{ij}=\pm 1,\pm\mathrm{\textit{i}}\}}\sigma_{ij}e^{-\beta\mathcal{H}_{ij}}\rangle_{G'}}{\langle\sum_{\{\sigma_{ij}=\pm 1,\pm\mathrm{\textit{i}}\}}e^{-\beta\mathcal{H}_{ij}}\big\rangle_{G'}}}.\\
\end{aligned}
\end{equation}
$\langle...\rangle_{G'}$ means the ensemble average over other parts of the graph that do not contain $\sigma_{ij}$. Now, we define $p\equiv\langle\sigma_{ij}\rangle$. 

As a mean-field approximation, we replace $\sigma_{jk}\sigma_{ki}$ with its ensemble average $q\equiv\langle\sigma_{jk}\sigma_{ki}\rangle$.  So, Eq. \ref{eq9} can be written as:
\begin{equation}\label{eq11}
\begin{aligned}
-\mathcal{H}_{ij}&\stackrel{\text{\textit{MF}}}\approx Re \left[\sigma_{ij} (N-2)q \right] +Im \left[\sigma_{ij} (N-2)q \right]+h\sigma_{ij}^2.
\end{aligned}
\end{equation} 
Since $p$ and $q$ are complex numbers, we name the real and imaginary part of them as follows:
\begin{equation}\label{eq12}
\begin{cases}
p_r= Re(p)\;\;;\;\;p_i= Im(p)\\
q_r= Re(q)\;\;;\;\;q_i=Im(q).
\end{cases}
\end{equation}
We calculate Eq. \ref{eq11} for different value of $\sigma_{ij}$:
\begin{equation}\label{eq13}
\begin{aligned}
& -\mathcal{H}_{ij}(\sigma_{ij}=+1)=(N-2)(q_r+q_i)+h\\
& -\mathcal{H}_{ij}(\sigma_{ij}=-1)=-(N-2)(q_r+q_i)+h\\
& -\mathcal{H}_{ij}(\sigma_{ij}=+i)=(N-2)(q_r-q_i)-h\\
& -\mathcal{H}_{ij}(\sigma_{ij}=-i)=-(N-2)(q_r+q_i)-h.\\
\end{aligned}
\end{equation}
By inserting the above relations in Eq. \ref{eq10}, we obtain:
\begin{equation}\label{eq14}
p(q_r, q_i; N, \beta, h)=
\frac{e^{\beta h}\sinh\Big(\beta(N-2)(q_r+q_i)\Big)+\mathrm{\textit{i}}\,e^{-\beta h}\sinh\Big(\beta(N-2)(q_r-q_i)\Big)}{e^{\beta h}\cosh\Big(\beta(N-2)(q_r+q_i)\Big)+e^{-\beta h}\cosh\Big(\beta(N-2)(q_r-q_i)\Big)}.
\end{equation}

\begin{figure}[H]
	\centering
	\includegraphics[scale=0.8]{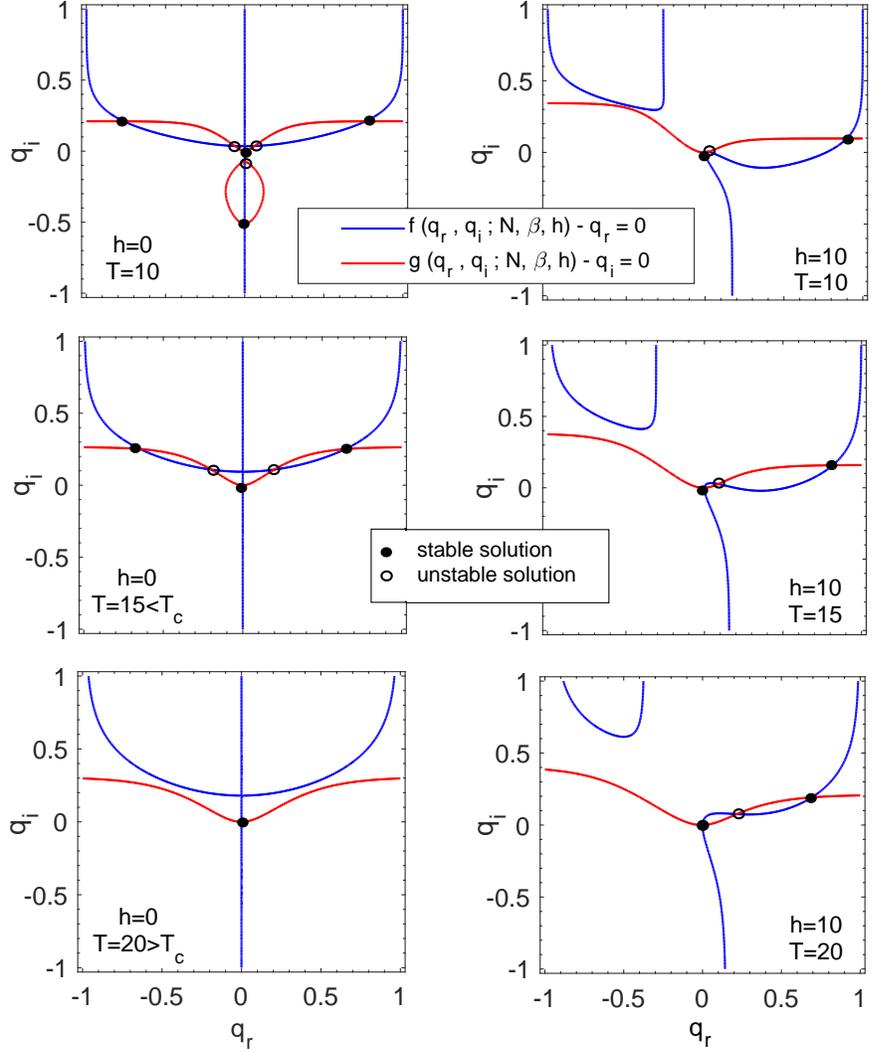}
	\caption{\textbf{Graphical representation of solutions of Eq. \ref{eq17} for a fully connected network with $N=50$ nodes at different temperatures $T=10, 15, 20$.} The left plots illustrate solutions in the absence of an external field $h=0$, while the right plots display solutions for a positive external field $h=10$. The blue and red curves indicate the solutions of $f(q_{r}, q_{i}; N,\beta,h)-q_r=0$ and $g(q_{r}, q_{i}; N,\beta,h)-q_i=0$, respectively. The intersections of blue and red curves are the simultaneous solutions of the two equations. Stable solutions are denoted by solid black points, while unstable solutions are denoted by hollow black points.} \label{Fig1}
\end{figure}

In the same way, we compute $q$:
\begin{equation}\label{eq15}
q(q_r, q_i; N, \beta, h)\equiv\langle\sigma_{jk}\sigma_{ki}\rangle=\frac{1}{Z}\sum_{{G}}\sigma_{jk}\sigma_{ki}e^{-\beta\mathcal{H}(G)}
=\frac{\big\langle\sum_{\{\sigma_{jk},\sigma_{ki}=\pm 1,\pm\mathrm{\textit{i}}\}}\sigma_{jk}\sigma_{ki}e^{-\beta\mathcal{H}_{\wedge_{ikj}}}\big\rangle_{G''}}{\big\langle\sum_{\{\sigma_{jk},\sigma_{ki}=\pm 1,\pm\mathrm{\textit{i}}\}}e^{-\beta\mathcal{H}_{\wedge_{ikj}}}\big\rangle_{G''}},
\end{equation}
where $\langle...\rangle_{G''}$ is ensemble average over all configurations which do not contain $\sigma_{jk}$ and $\sigma_{ki}$. After calculation we end up with:
\begin{equation}\label{eq16}
q\stackrel{\text{\textit{MF}}}{\approx}\frac{F(p,q;N , \beta, h)}{G(p,q; N, \beta, h)}.
\end{equation}
where details of the calculation and the explicit form of $F$ and $G$ have been provided in S1 Appendix. By calculating the real and imaginary part of $q$, we obtain two self-consistency equations: 
\begin{equation}\label{eq17}
\begin{cases}
q_r= Re[\frac{F(q,p;N,\beta,h)}{G(q,p;N,\beta,h)}]\equiv f(q_{r}, q_{i}; N,\beta,h)\\
\\
q_i=Im[\frac{F(q,p;N,\beta,h)]}{G(q,p;N,\beta,h)}]\equiv g(q_{r}, q_{i}; N,\beta,h).
\end{cases}
\end{equation}
Numerical solutions of both above equations are plotted separately in Fig. \ref{Fig1} within the allowed range of $q_r, q_i \in[-1,1]$. Simultaneous solutions of these two equations are the points where two curves cross each other. 

\begin{figure}[H]
	\centering
	\includegraphics[scale=0.8]{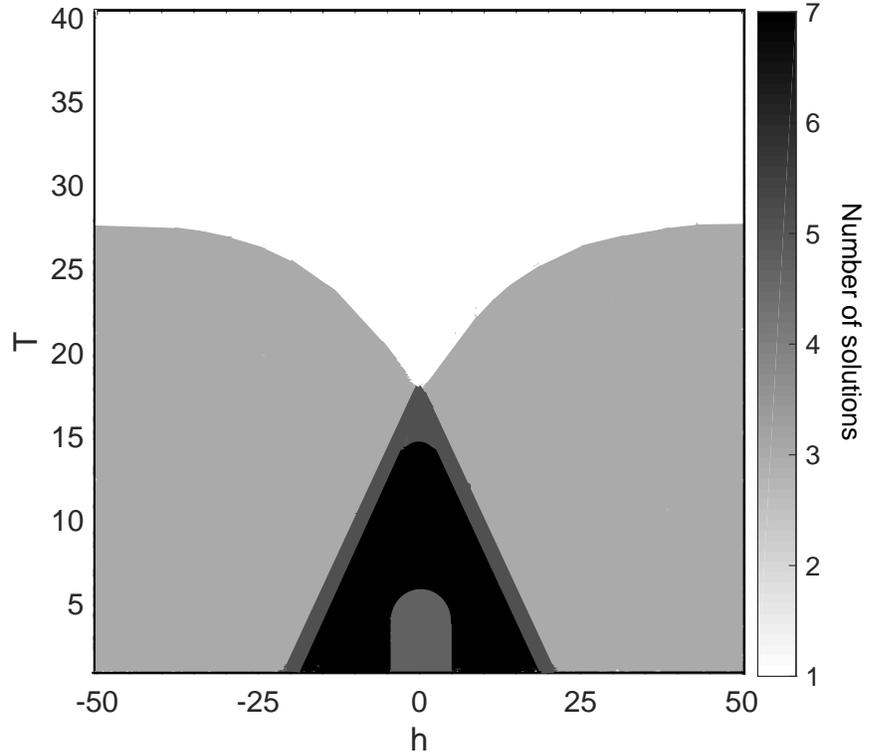}
	\caption{\textbf{Number of numeric solutions over the different values of external field $h$ and temperature $T$.} The plot shows how the number of solutions changes as $T$ and $h$ vary. Critical temperature in the absence of an external field ($h=0$) is $T_c=17.7$. } \label{Fig2}
\end{figure}

The number of solutions depends on the model’s free parameters i.e. field $h$ and temperature $T$. As depicted in Fig. \ref{Fig1} and Fig. \ref{Fig2}, we have more than one solution for some $h$ and $T$. Each solution is a pair of $(q_r, q_i)$. The stability and instability of these solutions are determined by analyzing the perturbation of solutions and their recursive update using Eq. \ref{eq17}; i.e. through analyzing whether the fixed points are attractive or repulsive \cite{masoumi2021mean, kargaran2020quartic}.

As the diagrams on the left column of Fig. \ref{Fig1} shows, in the absence of the external field $(h=0)$, the solutions exhibit symmetry with respect to positive and negative values of $q_r$. The solid black points represent stable solutions, while the hollow black points represent unstable solutions. On the right column of Fig. \ref{Fig1}, the external field has a positive value $(h=10)$, and as we expect the symmetry of the graphs disappears. As a result, there is no solution in the range $q_r<0$ for $T=10, 15, 20$ in $h=10$. There are three solutions in the region $q_r>=0$, so that two of them are stable and the other one is unstable. The behavior is reversed for negative external fields and as a result there is no solution in the range $q_r>0$.  Due to the perfect symmetry between positive and negative fields, only the diagrams for $h>0$ are plotted in Fig. \ref{Fig1}.

Fig. \ref{Fig2} illustrates phase diagram in $h$ and $T$ space. We achieve this figure by computing the number of solutions over the different values of external field and temperature.  As the figure shows, a phase transition occurs at the critical temperature $T_c =17.7$ for $h=0$. Above the transition temperature, there is one solution that corresponds to the unbalanced coexistence region, where both balanced and unbalanced triangles, as well as both real and imaginary links, are found in almost equal proportions in the network.

In the region below the critical temperature, multiple solutions are observed. In this region, one of the interests (real or imaginary) dominates the relations. So, the non-zero value for $q_r$ could be a stable solution. This shows the occurrence of a symmetry-breaking transition. 


So, to determine the dominant paradigm in the network, we calculate the mean value of generalized magnetization ($M$):
\begin{equation}\label{eq18}
\begin{aligned}
&\langle M \rangle=\langle\sigma_{ij}^2\rangle=\frac{1}{Z}\sum_{G}\sigma_{ij}^2 e^{-\beta\mathcal{H}(G)}=\\
&\frac{\sum_{\{\sigma\neq\sigma_{ij}\}}e^{-\beta\mathcal{H}'}\sum_{\{\sigma_{ij}=\pm 1,\pm\mathrm{\textit{i}}\}}\sigma_{ij}^2 e^{-\beta\mathcal{H}_{ij}}}{\sum_{\{\sigma\neq\sigma_{ij}\}}e^{-\beta\mathcal{H}'}\sum_{\{\sigma_{ij}=\pm 1,\pm\mathrm{\textit{i}}\}}e^{-\beta\mathcal{H}_{ij}}}=\\
&\frac{e^{2\beta h}\cosh\Big(\beta(N-2)(q_r+q_i)\Big)-\cosh\Big(\beta(N-2)(q_r-q_i)\Big)}{e^{2\beta h}\cosh\Big(\beta(N-2)(q_r+q_i)\Big)+\cosh\Big(\beta(N-2)(q_r-q_i)\Big)}.
\end{aligned}
\end{equation}

Ultimately, to determine whether a state is balanced or unbalanced, we compute $E_{\triangle}$. For calculating the mean value of ($E_{\triangle}$) we need to drive $r\equiv \langle \sigma_{ij}\sigma_{jk}\sigma_{ki} \rangle$ (see S1 Appendix):

\begin{equation}\label{eq19}
\begin{aligned}
&\langle E_{\triangle} \rangle=Re(r)+Im(r).
\end{aligned}
\end{equation}


By substituting mean-field solutions (each pair of $(q_r, q_i)$) into Eq. \ref{eq18} and Eq. \ref{eq19}, we can determine the value of $M$ and $E_{\triangle}$. For stable pairs of $(q_r, q_i)$, we obtain stable $M$ and $E_{\triangle}$. In the same way, unstable pairs of $(q_r, q_i)$ yield unstable values of $M$ and $E_{\triangle}$. The corresponding plots are shown in the result section.

\subsection{Simulation}
In addition to the mean-field solutions, we validate our results through simulations. We perform simulations on a fully connected network with $N=50$ nodes. The links of the network can take on one of four values: ${-1,+1,-i,+i}$. The network evolves via Monte-Carlo simulation.  In each update step, one link is randomly selected and with equal probability, it is converted to one of the three other types. This update will be accepted if the total energy of the network decreases. Otherwise, the conversion will be accepted with probability $p=exp(-\Delta E/kT)=exp(-\beta\Delta E)$, where $T$ indicates the network temperature and $\Delta E=E_2-E_1$ is the energy difference before and after the conversion. This process continues until the system reaches a relaxed state.

We examine the results of the simulations for three different initial conditions:
(1) a balanced network in which all links are $+1$.
(2) a balanced network in which all links are $-i$.
(3) an unbalanced network in which all links are randomly assigned values from $\pm1$, $\pm i$. 

\section{\label{sec4}Results and discussion}

\subsection{Response to the external field}
\begin{figure}[H]
	\centering
	\includegraphics[scale=0.8]{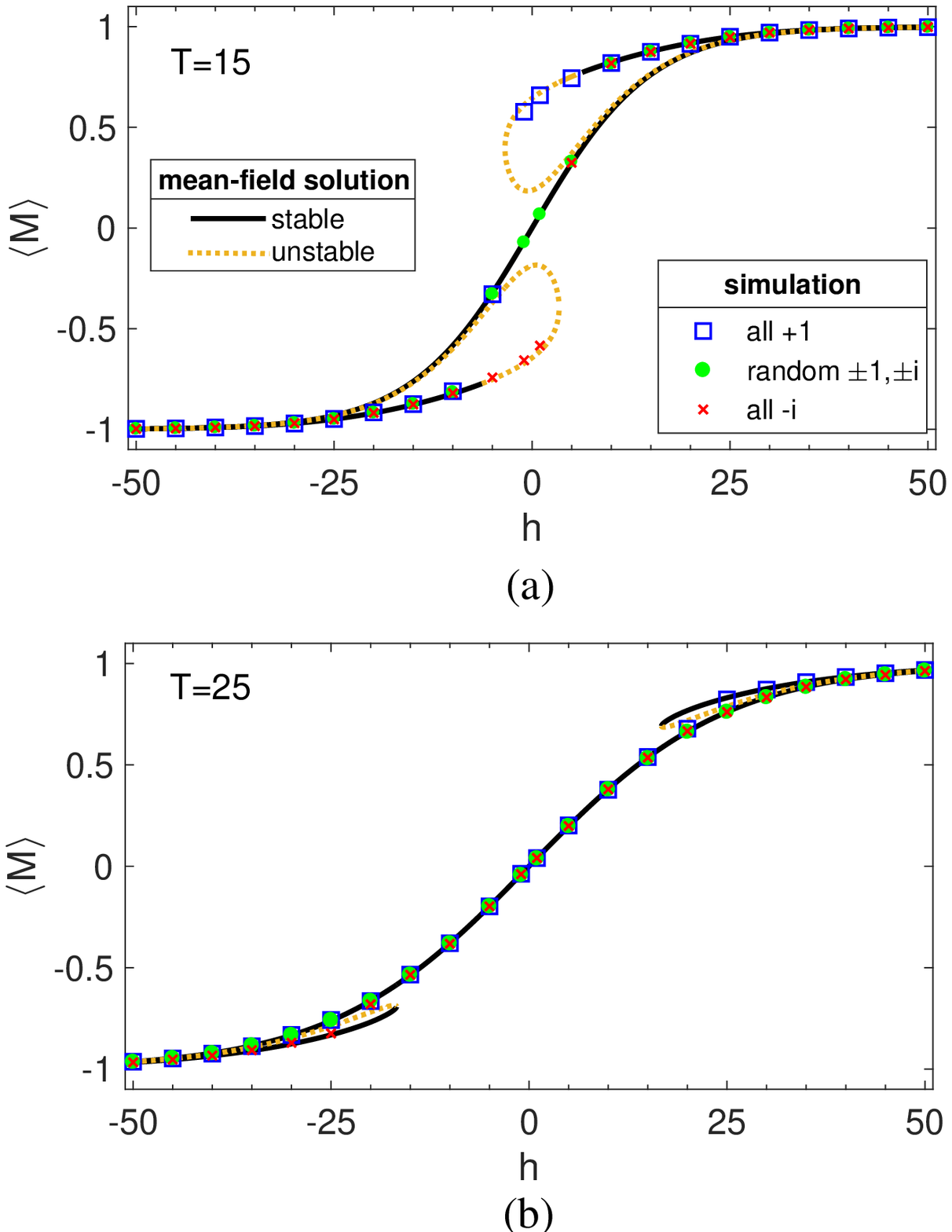}
	\caption{\textbf{Generalized magnetization  $M$ as a function of external field $h$ for temperatures below and above the critical temperature $T_c$: (a) $T=15<T_c$ , (b) $T=25>T_c$.}
These figures present a comparison between the mean-field solutions and the simulation results for different initial conditions. The initial conditions consist of two scenarios: a balanced network, where all links are either $+1$ or $-i$, and an unbalanced network with randomly assigned $\pm1$ and $\pm i$ links. The simulation results are  obtained from an ensemble of 100 realizations and are represented as the average. The solid black line and the dashed yellow line give the value of $M$ for stable and unstable mean-field solutions, respectively. } \label{Fig3}
\end{figure}

\begin{figure}[H]
	\centering
	\includegraphics[scale=0.8]{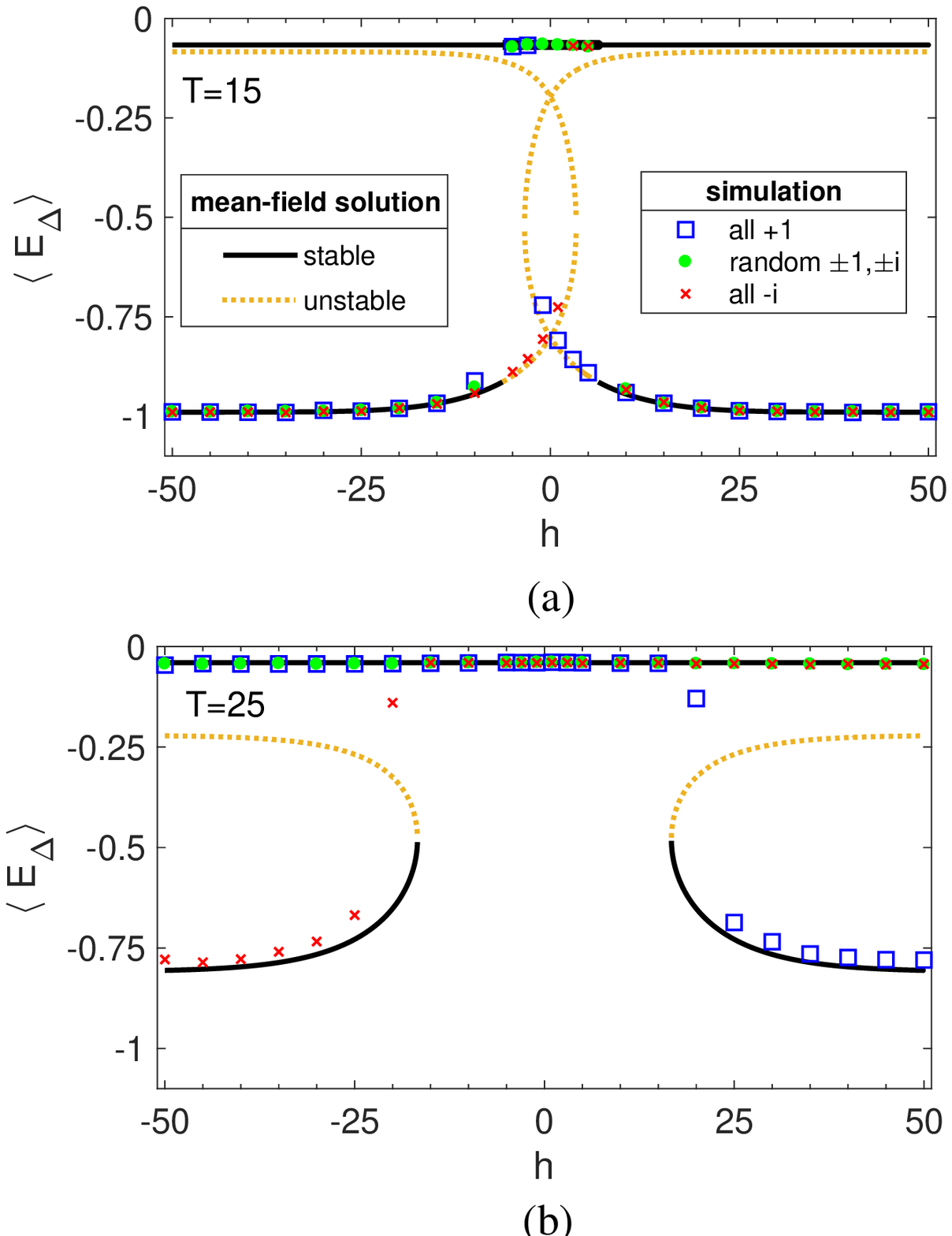}
      \caption{\textbf{Mean-triangle-energy $E_{\triangle}$ as a function of external field $h$ for temperatures below and above the critical temperature $T_c$: (a) $T=15<T_c$ , (b) $T=25>T_c$.}
These figures illustrate the comparison between the mean-field solutions and the simulation results for different initial conditions.  The initial conditions consist of two scenarios: a balanced network, where all links are either $+1$ or $-i$, and an unbalanced network with randomly assigned $\pm1$ and $\pm i$ links. The simulation results are are obtained from an ensemble of 100 realizations and are represented as the average. The solid black line and the dashed yellow line give the value of $E_{\triangle}$ for stable and unstable mean-field solutions, respectively. }\label{Fig4}
\end{figure}

Fig. \ref{Fig3} and Fig. \ref{Fig4} depict both analytical solutions and simulation results for the generalized magnetization $M$ and the mean-triangle-energy $E_{\triangle}$. As seen, there is a good agreement between simulation and the mean-field solutions, confirming the reliability of our approach. 
Fig. \ref{Fig3} represents the value of the generalized magnetization $M$  as a function of the external field for two temperatures: $T=15$ (below $T_c$) and $T=25$ (above $T_c$), while Fig. \ref{Fig4} represents the mean-triangle-energy $E_{\triangle}$ at the same temperatures. The simulation results are obtained by averaging over an ensemble of 100 realization. These plots clearly demonstrate that the values of macro variables are influenced by the initial conditions, which suggests the presence of hysteresis in the system. This observation aligns with the principles of the structural balance theory \cite{rabbani2019mean}.

In the absence of an external field, the Hamiltonian exhibits symmetry between both paradigms. However, the introduction of an external field breaks this symmetry and favors one paradigm over the other.
As depicted in Fig. \ref{Fig3} the non-zero value of the generalized magnetization is a consequence of the external field. In Fig. \ref{Fig3}(a) and Fig. \ref{Fig3}(b), we observe two distinct regimes characterized by temperature: the paramagnetic regime and the ferromagnetic regime. Above the critical temperature, the magnetization shows a linear increase with the field strength. However, upon removing the external field, the magnetization returns to zero, indicating the absence of any permanent magnetization, indicating the paramagnetic phase. In contrast, below the critical temperature, the network exhibits nonlinear behavior. Even after the field is removed, a residual magnetization is retained, indicating the presence of a ferromagnetic phase.  

However, in the competitive balance model, a value close to one for the generalized magnetization simply signifies the dominance of one paradigm in the system, while the energy curve reveals that tension still persists in the network or not.
In Fig. \ref{Fig4} we see that the level of energy is different for the two temperatures. For $T=15<T_c$ the value of mean-triangle-energy is close to $-1$ which corresponds to a balanced network. For $T=25$ which is above the critical temperature it is observed that while for strong external fields, the absolute value of magnetization is close to one, the level of mean-triangle-energy is close to zero. In magnetic systems, as the strength of the external field grows, the spins align, leading to a decrease in energy.   
Unlike the Ising model, in which a strong external field can break the symmetry and minimize the total energy in the paramagnetic phase, the competitive balance model in the paramagnetic phase, fails to minimize the mean-triangle-energy even when one paradigm dominates. In other words, in the competitive balance model, the dominance of one paradigm does not necessarily mean a reduction of tension.  Fig. \ref{Fig4} illustrates such differences clearly.

Overall, in the ferromagnetic regime, the network attains a state of balance, and the imposition of an external stimulus enables the network to achieve the desired paradigm. Notably, the presence of hysteresis in this regime indicates a resistance to paradigm shifts once dominance is established. As a result, sustained external stimulation is necessary until the network reaches the desired balanced state. Once achieved, the external field can be removed, and the desired paradigm will persist due to the phenomenon of hysteresis.
Conversely, the paramagnetic regime does not lead to network balance. External fields can provide temporary relief from unfavorable paradigms without completely eliminating tension. In this regime, the continuous imposition of the external field is crucial since removing the stimulation reverts the relationships to a binary choice between the two paradigms.

In summary, the simulation results reinforce the analytical solutions, highlighting the influence of initial conditions and the existence of a paramagnetic phase. Moreover, the findings suggest that the dominance of one paradigm in the competitive balance model does not always result in tension reduction, as evidenced by the mean-triangle-energy behavior.

\begin{figure}[H]
	\centering
       \includegraphics[scale=0.8]{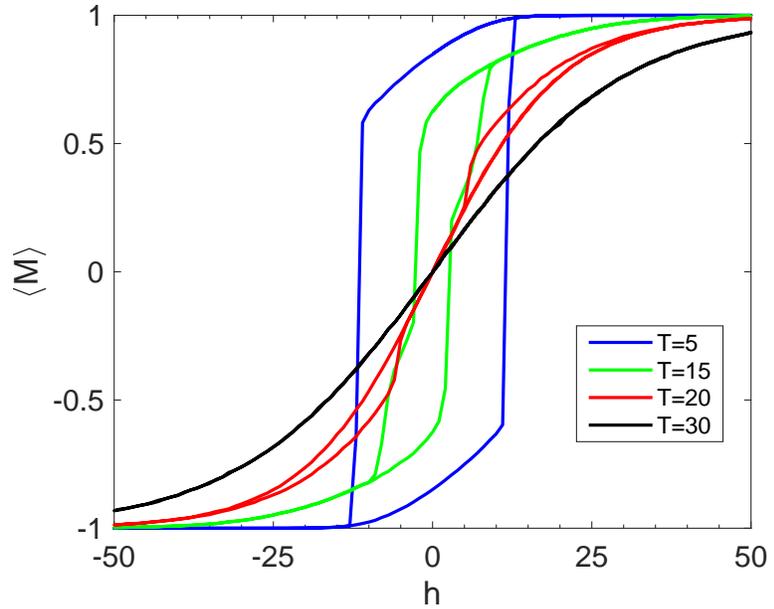}
	\caption{\textbf{Hysteresis loop for different temperatures $T$=5, 15, 20, 30.} Generalized magnetization M against magnetic field h. The size of the hysteresis loop decreases as the temperature increases.} \label{Fig5}
\end{figure}

{\bf Hysteresis-} Since the system has hysteresis, we aimed to figure out the hysteresis diagram. The result has been depicted in Fig. \ref{Fig5}. It is evident that the generalized magnetization $M$ does not have a single value for a given external field $h$ and its value depends on the system's history. This means that once communities form and a paradigm establishes dominance, it becomes challenging to change the situation and the system resists changes. 
Detecting hysteresis is important as it reveals the system's resistance to change and provides insights into the stability and memory effects of the system. Hysteresis in social networks indicates the persistence of certain beliefs, behaviors, or dominant paradigms even in the presence of external influences or attempts at change. The impact of temperature on hysteresis loops is depicted in Fig. \ref{Fig5}. It is observed that as the temperature grows, the hysteresis declines.  Since the phase transition is discrete, hysteresis vanishes above the critical temperature, where we call it a paramagnetic phase. This can be attributed to the discrete phase transition, where hysteresis vanishes above the critical temperature, referred to as the paramagnetic phase.

\begin{figure}[H]
	\centering
	\includegraphics[scale=0.8]{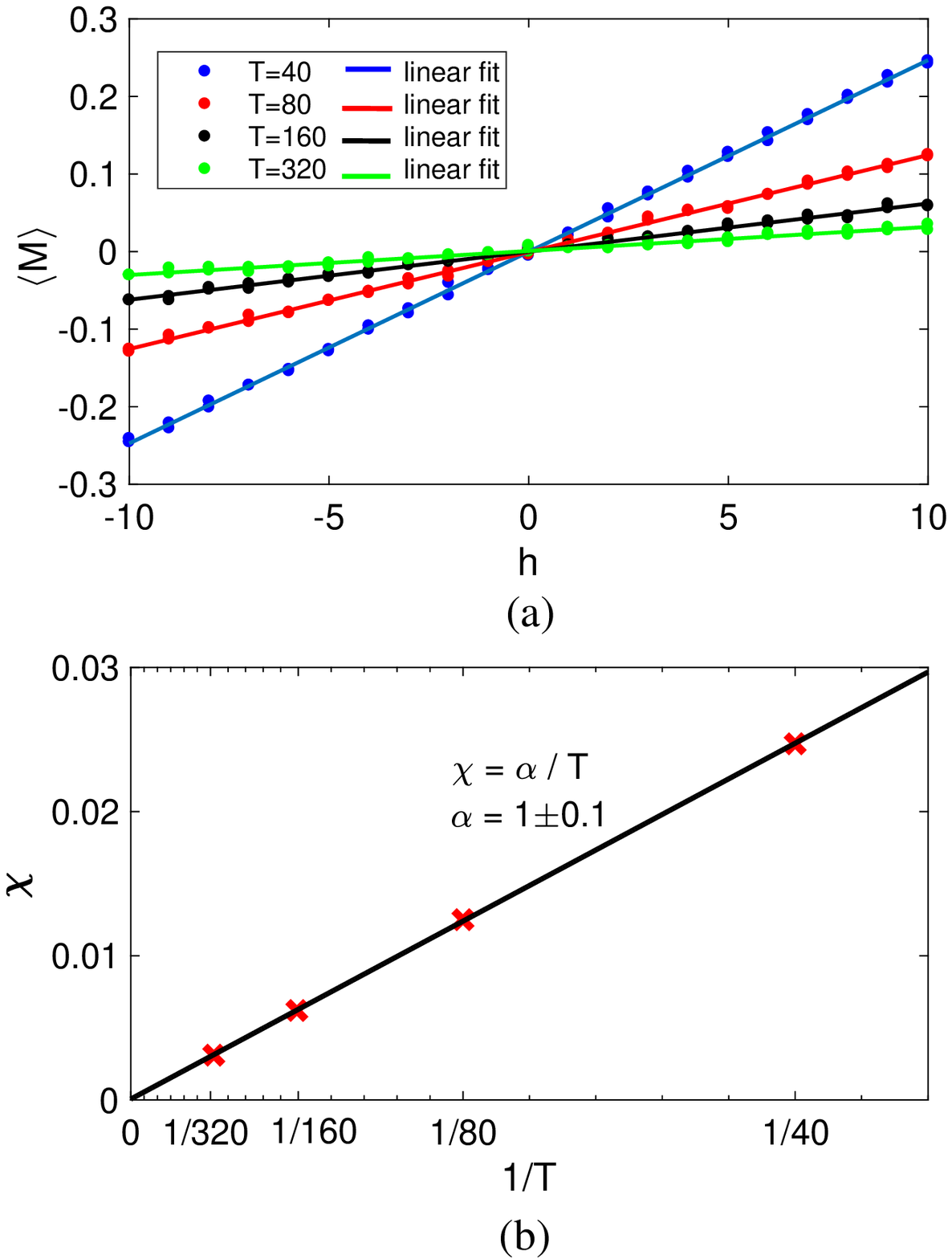}
	\caption{\textbf{Susceptibility $\chi$: The sensitivity of the generalized magnetization M to the external field.} \textbf{(a)} Generalized magnetization plotted against the small external fields $h=[-10,10]$, for temperatures far above the critical temperature i.e. $T_c \ll T=40, 80, 160, 320$;  the slope of the line corresponding to each temperature indicates susceptibility.  \textbf{(b)} Susceptibility plotted against the inverse of temperature, revealing the observation of Curie's law.} \label{Fig6}
\end{figure}

{\bf The Curie's law-} Another interest is investigating the impact of temperature on susceptibility, which measures the sensitivity of the generalized magnetization to the applied field $h$. The susceptibility is defined as $\mathcal{\chi}=\frac{\partial \langle M\rangle}{\partial h}$. Similarities with the magnetic systems raise interest in the study of Curie's law in the paramagnetic phase. Fig. \ref{Fig6} represents the result. As it can be seen, susceptibility has a linear relationship with
the inverse of temperature.
Detecting Curie's law in a model is important as it provides insights into the behavior of the system. By observing if the model's susceptibility aligns with Curie's law, we gain valuable information about the magnetic properties and behavior of the system being studied. This knowledge can aid in understanding the system's dynamics and predicting its response to external influences.

\subsection{Discussion}
To clear the concept, let's consider an application scenario in international relations, specifically focusing on the Middle Eastern countries. These countries often have religious-based relationships among themselves, while their relations with other countries are typically politically or economically motivated. In our model, we connect two countries via a real value link if the relation is political/economic, and in the religious case, we connect them via an imaginary link. It is important to note that relationship between countries can change over time. In this model, we specify the degree of stochastic behavior with the temperature parameter $T$.

Now, let’s suppose a large institution, organization, or government with its own reasons aims to influence and promote relations between countries. For example, it wants to support and expand relations with religious tendencies. It invests resources in advertising and other methods to achieve its goals. In order to add this external stimulation $h$, we can add a term to Hamiltonian as an external field.
The key point is to find the optimal value for stimulation. If the field is too weak, it may not be sufficient to shift the paradigm, and the network will not reach the desired balanced state. Consequently, the invested resources and efforts would be wasted. Conversely, if the field is too strong, exceeding the required amount, it becomes excessive and wasteful.

We also need to know how the system responds to external stimulation. Will the network align with the imposed field? Let's consider a situation where the mentioned institution, successfully establishes relations based on religious orientations. Now the question is, if the external field is removed, will the relations remain religious, or will the relations change as soon as the field is cut off? 
Our results show, depending on the level of stochastic behavior (temperature), there are two regimes: a ferromagnetic regime and a paramagnetic regime which are established below and above the critical temperature respectively.
In ferromagnetic regime, we have observed that once a paradigm dominates relations in a community, it then has hysteresis and thereby resists change. This can be observed in the long-standing conflicts and tensions in the Middle East, where despite being allies of the United States, countries such as Iraq, Israel, and Saudi Arabia have had ongoing conflicts for decades. The point is that the relationship between these countries and the US has economic/political origins, and the tension between them has a religious origin.

While the structural balance model provides a simple yet powerful framework for analyzing triangular relationships, it is necessary to incorporate heterogeneities in relations to capture real-world phenomena in greater detail.

\section{\label{sec6} Conclusion }

Heider's balance theory focused on achieving network balance when there is only one paradigm involved. However, in real-world societies, there can be multiple paradigms at play. Competitive balance theory expanded on this by considering the presence of more than one paradigm, but assumed equal competition between two paradigms. Although, in the real world, this equality is not always maintained and the symmetry is often broken by the application of external fields. In our study, we investigate the phase space of the system by incorporating the diversity of opinions and the level of stochastic behavior (temperature) when external forces are imposed. We find that a balanced state is achieved when a single paradigm prevails.  Additionally, we explore the cost of modifying a stable paradigm. When a paradigm has become stable in a system, the challenge lies in determining the cost of changing the current paradigm under external influence. When the level of stochastic behavior in the system is low, we observe that once a paradigm prevails, the system becomes resistant to change, exhibiting hysteresis behavior. Our findings are validated through both mean-field analysis and Monte Carlo simulations, confirming their consistency.

Overall, our research deepens our understanding of paradigm interactions, achieving network balance, and the role of external fields in real-world networks. Practically, our findings have important implications for understanding and managing paradigms within networks. By considering the level of stochastic behavior and the influence of external fields, individuals and organizations can optimize their efforts, reduce costs, and achieve long-term success in shaping desired paradigms.



\section*{Supporting information}

\hspace{0.5cm}{\bf S1 Appendix. Mean-Field solution for two-body and three body interactions.} Part A contains the mathematical calculations and explicit form for the two-body interactions ($q$). Part B elaborates on the detailed mathematical computations for the Mean-Field solution of three-body interactions ($\langle E_{\triangle} \rangle$).

\vspace{12pt}
{\bf S1 Table. Microsoft Excel database containing all the data obtained from the mean-field approximation method and simulation method used in this study. } Excel sheets in order: The sheet entitled “Number of numeric solutions” contains the number of simultaneous solutions of Eq. \ref{eq17} for a complete network with $N=50$ nodes over the different values of external field $h$ and temperature $T$. The sheet entitled “Mean-field solutions data” contains the simultaneous solutions of Eq. \ref{eq17} for temperatures $T=15$ and $T=25$ and various external fields $h$, while the stable and unstable solutions are separated.
The sheet entitled “Simulation data” contains the 6 simulated data sets that were generated using Monte-Carlo simulation: 3 different initial conditions for 2 temperatures $T=15$ and $T=25$ over a wide range of external fields. 

\section*{Appendix}
\appendix
\section{Mean-Field solution for two-body term $q$}\label{A}
At first step, we separate all terms that contain $\sigma_{jk}$ or $\sigma_{ki}$:
\small
\begin{equation}
\mathcal{H}=\mathcal{H}_{\wedge_{jki}}+\mathcal{H}''
\end{equation}
\begin{equation}
\begin{aligned}
-\mathcal{H}_{\wedge_{jki}}&=Re \left[\sigma_{jk}\sum_{\ell\neq i,j,k}\sigma_{j\ell}\sigma_{\ell k}\right]+Im \left[\sigma_{jk}\sum_{\ell\neq i,j,k}\sigma_{j\ell}\sigma_{\ell k}\right]+Re \left[\sigma_{ki}\sum_{\ell\neq i,j,k}\sigma_{k\ell}\sigma_{\ell i}\right]\\
&\quad+Im \left[\sigma_{ki}\sum_{\ell\neq i,j,k}\sigma_{k\ell}\sigma_{\ell i}\right]
+Re \left(\sigma_{ij}\sigma_{jk}\sigma_{ki}\right)+Im \left(\sigma_{ij}\sigma_{jk}\sigma_{ki}\right) +h(\sigma_{jk}^2+\sigma_{ki}^2)\\
&\stackrel{\text{\textit{MF}}}\approx Re \left[\sigma_{jk}(N-3)q\right]+Im \left[\sigma_{jk}(N-3)q\right]+Re \left[\sigma_{ki}(N-3)q\right]+Im \left[\sigma_{ki}(N-3)q \right]\\
&\;\;\;\;\;\;+Re \left(\sigma_{ij}\sigma_{jk}\sigma_{ki}\right)+Im \left(\sigma_{ij}\sigma_{jk}\sigma_{ki}\right) +h(\sigma_{jk}^2+\sigma_{ki}^2)\\
\end{aligned}
\end{equation}
\normalsize
Different modes that these two links can take are
\begin{equation}
\begin{aligned}
& -\mathcal{H}_{\wedge_{ikj}}(\sigma_{jk}=+1, \sigma_{ki}=+1)=2(N-3)(q_r+q_i)+p_r+p_i +2h\\
&-\mathcal{H}_{\wedge_{ikj}}(\sigma_{jk}=-1, \sigma_{ki}=-1)=-2(N-3)(q_r+q_i)+p_r+p_i+2h \\
&-\mathcal{H}_{\wedge_{ikj}}(\sigma_{jk}=+1, \sigma_{ki}=-1)=-(p_r+p_i)+2h \; \to \; (\times2) \\
&-\mathcal{H}_{\wedge_{ikj}}(\sigma_{jk}=+i, \sigma_{ki}=+i)=2(N-3)(q_r-q_i)-(p_r+p_i)-2h \\
&-\mathcal{H}_{\wedge_{ikj}}(\sigma_{jk}=-i, \sigma_{ki}=-i)=2(N-3)(-q_r+q_i)-(p_r+p_i)-2h \\
&-\mathcal{H}_{\wedge_{ikj}}(\sigma_{jk}=+i, \sigma_{ki}=-i)=p_r+p_i-2h \; \to \; (\times2) \\
&-\mathcal{H}_{\wedge_{ikj}}(\sigma_{jk}=+1, \sigma_{ki}=+i)=2(N-3)q_r+p_r-p_i \; \to \; (\times2) \\
&-\mathcal{H}_{\wedge_{ikj}}(\sigma_{jk}=+1, \sigma_{ki}=-i)=2(N-3)q_i-p_r+p_i \; \to \; (\times2) \\
&-\mathcal{H}_{\wedge_{ikj}}(\sigma_{jk}=-1, \sigma_{ki}=+i)=-2(N-3)q_i-p_r+p_i \; \to \; (\times2) \\
&-\mathcal{H}_{\wedge_{ikj}}(\sigma_{jk}=-1, \sigma_{ki}=-i)=-2(N-3)q_r+p_r-p_i \; \to \; (\times2) \\
\end{aligned}
\end{equation}
By substituting above relations into Eq. 15, we abtain following equation:
\begin{equation}
\langle\sigma_{jk}\sigma_{ki}\rangle\stackrel{\text{\textit{MF}}}{\approx}\frac{F(p,q;N , \beta, h)}{G(p,q; N, \beta, h)},
\end{equation}
where
\begin{equation}
\begin{aligned}
F(p,q; N,\beta, h)&=
e^{\beta[2(N-3)(q_r+q_i)+p_r+p_i+2h]}
+e^{\beta[-2(N-3)(q_r+q_i)+p_r+p_i+2h]}
-2e^{\beta[-(p_r+p_i)+2h]}\\
&-e^{\beta[2(N-3)(q_r-q_i)-(p_r+p_i)-2h]}
-e^{\beta[2(N-3)(-q_r+q_i)-(p_r+p_i)-2h]}
+ 2e^{\beta[p_r+p_i-2h]}\\
&+2\mathrm{\textit{i}}\,e^{\beta[2(N-3)q_r+p_r-p_i]}
-2\mathrm{\textit{i}}\,e^{\beta[2(N-3)q_i-p_r+p_i]}
-2\mathrm{\textit{i}}\,e^{\beta[-2(N-3)q_i-p_r+p_i]}\\
&+2\mathrm{\textit{i}}\,e^{\beta[-2(N-3)q_r+p_r-p_i]},
\end{aligned}
\end{equation}
\begin{equation}
\begin{aligned}
G(p,q; N, \beta, h)&=
e^{\beta[2(N-3)(q_r+q_i)+p_r+p_i+2h]}
+e^{\beta[-2(N-3)(q_r+q_i)+p_r+p_i+2h]}
+2e^{\beta[-(p_r+p_i)+2h]}\\
&+e^{\beta[2(N-3)(q_r-q_i)-(p_r+p_i)-2h]}
+e^{\beta[2(N-3)(-q_r+q_i)-(p_r+p_i)-2h]}
+ 2e^{\beta[p_r+p_i-2h]}\\
&+2e^{\beta[2(N-3)q_r+p_r-p_i]}
+2e^{\beta[2(N-3)q_i-p_r+p_i]}
+2e^{\beta[-2(N-3)q_i-p_r+p_i]}\\
&+2e^{\beta[-2(N-3)q_r+p_r-p_i]}.
\\
\end{aligned}
\end{equation}
\newpage

\section{Mean-Field solution for three body interactions}\label{B}
Similar to the previous sections, at first, we must separate the sentences that contain links $\sigma_{ij}$, $\sigma_{jk}$, $\sigma_{ki}$:
\begin{equation}
\mathcal{H}=\mathcal{H}_{\triangle_{ijk}}+\mathcal{H}'''
\end{equation}

\begin{equation}
\begin{aligned}
&r\equiv\langle\sigma_{ij}\sigma_{jk}\sigma_{ki}\rangle=\frac{1}{Z}\sum_{{G}}\sigma_{ij}\sigma_{jk}\sigma_{ki}e^{-\beta\mathcal{H}(G)}\\
&=\frac{\sum_{\{\sigma\neq\sigma_{ij},\sigma_{jk},\sigma_{ki}\}}e^{-\beta\mathcal{H}'''}\sum_{\{\sigma_{ij},\sigma_{jk},\sigma_{ki}=\pm 1,\pm\mathrm{\textit{i}}\}}\sigma_{ij}\sigma_{jk}\sigma_{ki}e^{-\beta\mathcal{H}_{\triangle_{ikj}}}}{\sum_{\{\sigma\neq\sigma_{ij},\sigma_{jk},\sigma_{ki}\}}e^{-\beta\mathcal{H}'''}\sum_{\{\sigma_{ij},\sigma_{jk},\sigma_{ki}=\pm 1,\pm\mathrm{\textit{i}}\}}e^{-\beta\mathcal{H}_{\triangle_{ikj}}}}\\
\end{aligned}
\end{equation}
\begin{equation}
\begin{aligned}
-\mathcal{H}_{\triangle_{ijk}}&=Re \left[\sigma_{ij}\sum_{\ell\neq i,j,k}\sigma_{i\ell}\sigma_{\ell j}\right]+Im \left[\sigma_{ij}\sum_{\ell\neq i,j,k}\sigma_{i\ell}\sigma_{\ell j}\right]
+Re \left[\sigma_{jk}\sum_{\ell\neq i,j,k}\sigma_{j\ell}\sigma_{\ell k}\right]\\
&\quad+Im \left[\sigma_{jk}\sum_{\ell\neq i,j,k}\sigma_{j\ell}\sigma_{\ell k}\right]+Re \left[\sigma_{ki}\sum_{\ell\neq i,j,k}\sigma_{k\ell}\sigma_{\ell i}\right]+Im \left[\sigma_{ki}\sum_{\ell\neq i,j,k}\sigma_{k\ell}\sigma_{\ell i}\right]\\
&\quad+Re \left(\sigma_{ij}\sigma_{jk}\sigma_{ki}\right)+Im \left(\sigma_{ij}\sigma_{jk}\sigma_{ki}\right)
+h(\sigma_{ij}^2+\sigma_{jk}^2+\sigma_{ki}^2)\\
&\stackrel{\text{\textit{MF}}}\approx Re \left[\sigma_{ij}(N-3)q\right]+Im \left[\sigma_{ij}(N-3)q\right]
+Re \left[\sigma_{jk}(N-3)q\right]+Im \left[\sigma_{jk}(N-3)q \right]\\
&\quad+Re \left[\sigma_{ki}(N-3)q\right]+Im \left[\sigma_{ki}(N-3)q \right]
+Re \left(\sigma_{ij}\sigma_{jk}\sigma_{ki}\right)+Im \left(\sigma_{ij}\sigma_{jk}\sigma_{ki}\right)\\ 
&\quad+h(\sigma_{ij}^2+\sigma_{jk}^2+\sigma_{ki}^2)
\end{aligned}
\end{equation}
\begin{equation}
\begin{aligned}
r_r=Re(r), \;\;\;\;\;\;\; r_i=Im(r),
\end{aligned}
\end{equation}
\begin{equation}
\begin{aligned}
\langle E_{\triangle}\rangle=r_r+r_i.
\end{aligned}
\end{equation}
So we have:
\begin{equation}
\langle E_{\triangle} \rangle\stackrel{\text{\textit{MF}}}{\approx}\frac{V(q;N , \beta, h)}{W(q; N, \beta, h)},
\end{equation}
where:
\begin{equation}
\begin{aligned}
V(q; N,\beta, h)&=
-3e^{\beta[-(N-3)(3q_r-q_i)+1-h]}
-e^{\beta[-3(N-3)(q_r-q_i)+1-3h]}
-3e^{\beta[-(N-3)(q_r+3q_i)+1+h]}\\
&\quad-3e^{\beta[-(N-3)(q_r+q_i)+1+3h]}
-6e^{\beta[-(N-3)(q_r-q_i)+1+h]}
-3e^{\beta[(N-3)(q_r-3q_i)+1-h]}\\
&\quad+3e^{-\beta[(N-3)(q_r-3q_i)+1+h]}
-3e^{\beta[(N-3)(q_r-q_i)+1-3h]}
-6e^{\beta[(N-3)(q_r+q_i)+1-h]}\\
&\quad+6e^{-\beta[(N-3)(q_r+q_i)+1+h]}
-3e^{\beta[(N-3)(3q_r+q_i)+1+h]}
-e^{\beta[3(N-3)(q_r+q_i)+1+3h]}\\
&\quad+6e^{\beta[(N-3)(q_r-q_i)-1+h]}
+3e^{\beta[(N-3)(q_r+q_i)-1+3h]}
+3e^{-\beta[(N-3)(q_r+3q_i)-1+h]}\\
&\quad+e^{\beta[3(N-3)(q_r-q_i)-1-3h]}
+3e^{\beta[-(N-3)(3q_r+q_i)-1+h]}
+3e^{-\beta[-(N-3)(3q_r-q_i)+1+h]}\\
&\quad+e^{\beta[-3(N-3)(q_r+q_i)-1+3h]}
+3e^{-\beta[-(N-3)(q_r-q_i)-1-3h]}
\end{aligned}
\end{equation}
\begin{equation}
\begin{aligned}
W(q; N, \beta, h)&=
3e^{\beta[-(N-3)(3q_r-q_i)+1-h]}
+e^{\beta[-3(N-3)(q_r-q_i)+1-3h]}
+3e^{\beta[-(N-3)(q_r+3q_i)+1+h]}\\
&+3e^{\beta[-(N-3)(q_r+q_i)+1+3h]}
+6e^{\beta[-(N-3)(q_r-q_i)+1+h]}
+3e^{\beta[(N-3)(q_r-3q_i)+1-h]}\\
&+3e^{-\beta[(N-3)(q_r-3q_i)+1+h]}
+3e^{\beta[(N-3)(q_r-q_i)+1-3h]}
+6e^{\beta[(N-3)(q_r+q_i)+1-h]}\\
&+6e^{-\beta[(N-3)(q_r+q_i)+1+h]}
+3e^{\beta[(N-3)(3q_r+q_i)+1+h]}
+e^{\beta[3(N-3)(q_r+q_i)+1+3h]}\\
&+6e^{\beta[(N-3)(q_r-q_i)-1+h]}
+3e^{\beta[(N-3)(q_r+q_i)-1+3h]}
+3e^{-\beta[(N-3)(q_r+3q_i)-1+h]}\\
&+e^{\beta[3(N-3)(q_r-q_i)-1-3h]}
+3e^{\beta[-(N-3)(3q_r+q_i)-1+h]}
+3e^{-\beta[-(N-3)(3q_r-q_i)+1+h]}\\
&+e^{\beta[-3(N-3)(q_r+q_i)-1+3h]}
+3e^{-\beta[-(N-3)(q_r-q_i)-1-3h]}
\end{aligned}
\end{equation}

\end{document}